\documentstyle[aps,twocolumn]{revtex}
\input epsf.tex

\newcommand{\be}{\begin{equation}}
\newcommand{\ee}{\end{equation}}
\newcommand{\ba}{\begin{eqnarray}}
\newcommand{\ea}{\end{eqnarray}}
\newcommand{\ban}{\begin{eqnarray*}}
\newcommand{\ean}{\end{eqnarray*}}

\newcommand{\sandwich}[3]{\mbox{$ \langle #1 | #2 | #3 \rangle $}}
\newcommand{\ket}[1]{\mbox{$ | #1 \rangle $}}

\newcommand{\si}{\sigma}

\newcommand{\one}{\leavevmode\hbox{\small1\normalsize\kern-.33em1}}

\newcommand{\Tr}{\mbox{Tr}}

\newcommand{\moy}[1]{\langle #1 \rangle}

\begin{document}

\title{Quantum Communication between N-partners and Bell's Inequalities}

\author{ Valerio Scarani\thanks{valerio.scarani@physics.unige.ch},
Nicolas Gisin \\
Group of Applied Physics, University of Geneva\\
20, rue de l'Ecole-de-M\'edecine, CH-1211 Geneva 4, Switzerland\\
 }
%\date{}
\maketitle

\begin{abstract}
We consider a family of quantum communication protocols involving
$N$ partners. We demonstrate the existence of a link between the
security of these protocols against individual attacks by the
eavesdropper, and the violation of some Bell's inequalities,
generalizing the link that was noticed some years ago for
two-partners quantum cryptography. The arguments are independent
of the local hidden variable debate.
\end{abstract}

%\vspace{1cm}

%Keywords:

\vspace{1cm}

Historically, entanglement was essentially a source of controversy
on the foundations of quantum mechanics, as illustrated by the
lively debate about the local hidden variable program and Bell's
inequality \cite{BellSpeakable}. Today, it is widely recognized
that entanglement is a resource from which tasks can be achieved
that are classically impossible, as illustrated by many quantum
information protocols \cite{PhysQInfo}. Among these protocols,
quantum cryptography --- better described as quantum key
distribution (QKD) --- is the one that has almost reached the
level of application \cite{revue}. In this work, we study the link
between the security of quantum communication protocols and the
violation of Bell's inequalities. Previous works
\cite{bruno,fuchs} pointed out such a link in QKD between two
partners Alice and Bob. We begin by reviewing these results, that
will help to clarify the initial intuition and the motivation for
the present work.

Consider the following QKD setup \cite{ekert}. Alice prepares an
EPR state, say $\ket{\Phi^{+}_z}=
\frac{1}{\sqrt{2}}\left(\ket{00}+\ket{11}\right)$, where we write
$\ket{0}$ and $\ket{1}$ for the eigenstates of $\si_z$. She keeps
one qubit and sends the other one to Bob. Alice and Bob measure
either $\si_x$ or $\si_y$, then publicly communicate the choice of
the measurement basis. Whenever they have used the same basis,
their results are perfectly correlated, and they can establish a
key. This protocol is equivalent to the BB84 protocol \cite{bb84}.
Its distinguishing feature is the fact that the bits are encoded
into orthogonal states belonging to two conjugated bases.

To study the security of the protocol, consider an eavesdropper
(Eve) that acts on the quantum channel linking Alice to Bob,
trying to get some information but inevitably introducing
perturbations. To establish a key in spite of these perturbations,
A and B can run a one-way protocol called {\em error correction
and privacy amplification} if and only if \cite{csi} \be
I(A:B)\,>\,\min\,[I(A:E),I(B:E)] \label{condsec} \ee where
$I(A:B)=H(A)+H(B)-H(AB)$, $H$ the Shannon entropy, is called {\em
mutual information}. In the following, we shall consider
(\ref{condsec}) as the condition for security, although it is
known that a secret key can be established under less restrictive
conditions by using two-way communication \cite{wolf}.

In our context, Eve's best attack is defined as the attack that
maximizes $I(A:E)$ for a fixed $I(A:B)$. The best attack is not
known in all generality \cite{shor}; but it is, if we suppose that
Eve performs an {\em individual attack}, that is, that she makes
only measurements on individual qubits \cite{fuchs}. Moreover, it
is also known that Eve can perform the best individual attack by
using a single qubit as resource \cite{niu}, by implementing the
following unitary transformation affecting her and Bob's qubits:
\be
\begin{array}{lll}
U_{BE}\ket{00}&=&\ket{00}\\
U_{BE}\ket{10}&=&\cos\phi\ket{10}\,+\,\sin\phi\ket{01}
\end{array}\,.
\label{isometry} \ee Here, $\ket{00}$ etc. are shorthand for
$\ket{0}_B\otimes\ket{0}_E$ etc. (by convention, we supposed that
Eve prepares her qubits in the state $\ket{0}$), and
$\phi\in [0,\frac{\pi}{2}]$ characterizes the strength of Eve's
attack. Thus, after eavesdropping the system of three qubits is in
the state $\ket{\Psi_{ABE}}= \frac{1}{\sqrt{2}} (\ket{0}_A\otimes
U_{BE}\ket{00}+ \ket{1}_A\otimes U_{BE}\ket{10})$. Note that the
roles of B and E are symmetric under the exchange of $\phi$ with
$\frac{\pi}{2}-\phi$. The mutual information between any two
partners can be calculated explicitly \cite{full}: condition (\ref{condsec})
for security is fulfilled if and only if $\phi<\frac{\pi}{4}$.

\begin{figure}
\begin{center}
\epsfxsize=7cm \epsfbox{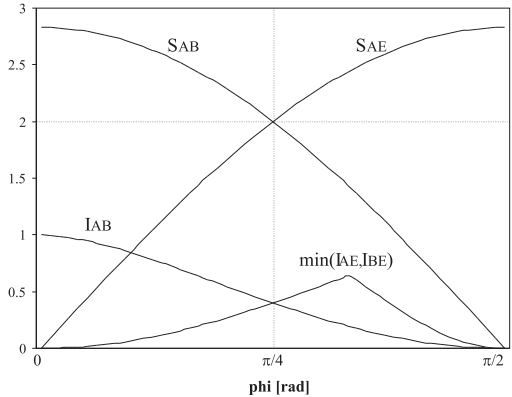} \caption{The link between
violation of Bell's inequality and the security condition
(\ref{condsec}), in the case of the two-partners QKD with best
individual attack by Eve.} \label{figqkd}
\end{center}
\end{figure}

As we said above, there is a remarkable link between the security
of the BB84 protocol against individual attacks and the violation
of Bell's inequalities. For any set of four unit vectors
$\underline{a}=\{\vec{a}_1,\vec{a}\,'_1,
\vec{a}_2,\vec{a}\,'_2\}$, let's define the two-qubit Bell
operator \ba B_2(\underline{a})&=&
\left(\sigma_{a_1}+\si_{a'_1}\right)\otimes\si_{a_2} +
\left(\sigma_{a_1}-\si_{a'_1}\right)\otimes\si_{a_2'}
\label{eqchsh}\ea with $\sigma_a=\vec{a}\cdot\vec{\si}$. The CHSH
inequality \cite{chsh} reads
$S_2=\max_{\underline{a}}\Tr(\rho\,B_2(\underline{a})) \leq 2$,
while the maximal value allowed by QM is $S_2=2\sqrt{2}$
\cite{cirelson}. The CHSH inequality is optimal, in the sense it
is violated if and only if the statistics of the results cannot be
accounted for by local hidden variables (lhv) \cite{bell}. The
Horodecki criterion \cite{horo} allows an explicit calculation of
$S$ for each two-qubit state obtained from $\ket{\Psi_{ABE}}$ by
tracing out the third qubit. We find that the pair B-E never
violates the inequality, while \ba S_{AB}=2\sqrt{2}\cos\phi
&\;,\;& S_{AE}=2\sqrt{2}\sin\phi\,.\ea Then obviously $S_{AB}>2$
if and only if $S_{AE}<2$: the inequality is violated by the pair
A-B if and only if it is not violated by the pair A-E. In
conclusion, for the QKD protocol that we consider, (\ref{condsec})
holds if and only if $S_{AB}>2>S_{AE}$ (fig. \ref{figqkd}).

The previous paragraphs summarize the present knowledge about the
link between security and Bell's inequalities. In the following,
we shall generalize this link for QKD protocols involving an
arbitrary number of partners. But before turning to this, let's
address the following purely algebraic problem, which is naturally
related to this discussion. Consider three partners A,B and C
(here there is no more reason to single out an Eve), each
possessing a qubit. Are there pure or mixed states of the three
qubit system such that more than one pair can violate
the CHSH inequality? The answer to this question is negative:\\
{\bf Theorem 1:}
{\em Let $\rho$ be a three-qubit state, and
$\rho_{AB}$, $\rho_{BC}$ and $\rho_{AC}$ be the two-qubit states
obtained from $\rho$ by tracing out one of the qubits. If one can
find four unit vectors $\vec{a},\vec{a}\,',\vec{b}, \vec{b}\,'$
such that $\Tr(B_{2}\,\rho_{AB})>2$, then for all choice of four
unit vectors $\Tr(B_{2}\,\rho_{BC})<2$ and
$\Tr(B_{2}\,\rho_{AC})<2$.}\\
We present a proof inspired by Cirel'son's proof that the maximal
violation of CHSH allowed by quantum mechanics is $2\sqrt{2}$
\cite{cirelson}. Let's define the operator \be V\,=\,
B_{AB}(\vec{a},\vec{a}\,',\vec{b}, \vec{b}\,')\,+\,
B_{AC}(\vec{A},\vec{A}\,',\vec{c}, \vec{c}\,')\,.\ee where
$B_{AB}=B_2\otimes\one_C$, and similarly for $B_{AC}$. Using
$\si_a\si_{a'}=(\vec{a}\cdot\vec{a}\,')\one+i\si_{a\wedge a'}$,
lengthy but standard algebra leads to $\left(\frac{V^2}{4}-
2\one\right)^2=f\one$, where $f$ is a function of the unit
vectors that satisfies $0\leq f\leq 4$ \cite{notef}. This entails
$\left|\moy{V}_{\rho}\right|\leq 4$, that is
$\mbox{max}\left|\moy{B_{AB}+B_{AC}}_{\rho}\right|\leq 4$, where
the maximum is taken over the eight unit vectors that define $V$.
But due to the symmetry $B_{AB}(\vec{a},\vec{a}\,',-\vec{b},
-\vec{b}\,')\,=\,-\,B_{AB}(\vec{a},\vec{a}\,',\vec{b},
\vec{b}\,')$, it holds that
$\mbox{max}\left|\moy{B_{AB}+B_{AC}}_{\rho}\right|=
\mbox{max}\left|\moy{B_{AB}}_{\rho}\right|
+\mbox{max}\left|\moy{B_{AC}}_{\rho}\right|= S_{AB}+S_{AC}$. In
conclusion, $S_{AB}+S_{AC}\leq 4$ for all $\rho$ and for all
choice of unit vectors. This proves the theorem.

Two remarks: (i) It is easy to imagine experimental protocols in
which, for suitable states, both pairs A-B and A-C end up with a
violation of the inequality: e.g., a pair can analyze their data
conditioning on the results of the third partner, if they know
this result through classical communication; or, a pair may apply
a filtering procedure \cite{gisin}. (ii) There are states
depending on one or more parameters such that one can "shift" the
violation from one pair to another by varying the parameters: the
state introduced above in the context of QKD, in particular, is
such that $S_{AB}(\phi)>2$ if and only if $S_{AC}(\phi)<2$
\cite{note1}.

We explore now the generalization of the link between Bell's
inequalities and security to QKD protocols involving more than two
partners. The protocols that we consider are characterized by the
fact that the sender distributes the key between several partners,
in such a way that {\em all partners must collaborate} to retrieve
the key. We call these protocols {\em N-partners quantum secret
sharing} (N-QSS) \cite{qss}. For simplicity, we discuss in detail
the protocol 3-QSS involving three partners, and discuss later how
this generalizes to an arbitrary number of partners. Without
eavesdropping, 3-QSS works as follows. Alice prepares the 3-qubit
GHZ state $\frac{1}{\sqrt{2}}(\ket{000}+\ket{111})$; she keeps one
qubit and sends the others to her two partners Bob and Charlie.
The three of them measure $\si_x$ or $\si_y$; the GHZ state is
such that $\moy{\si_x\otimes\si_x\otimes\si_x}=
-\moy{\si_x\otimes\si_y\otimes\si_y}=-\moy{\si_y\otimes\si_x\otimes\si_y}
=-\moy{\si_y\otimes\si_y\otimes\si_x}=1$, and the other
expectation values vanish. Then A, B and C publicly announce the
bases they used, and keep only those measurements when all
measured $\si_x$, or when one measured $\si_x$ and the others
$\si_y$. It is easy to see that each partner alone has no
information on the key of any other partner, but if two partners
collaborate then they have all the information about the key of
the third partner. Therefore the meaningful information measure is
the information that B and C together have on A's sequence of
bits, that is $I(A:BC)=H(A)-H(A|BC)=1-H(A|BC)$. In the absence of
eavesdropping, $H(A|BC)=0$.

Two eavesdropping scenarios can be imagined:\\
{\em Scenario 1:} An external Eve tries to eavesdrop on both channels A-B and A-C.
We still restrict to attacks that are "individual" in the sense that each pair of qubits is
attacked independently from all the other pairs; but we allow
coherent measurements on the two qubits of each pair. \\
{\em Scenario 2:} Charlie is dishonest: he would like to retrieve
the key alone, against the will of Alice who would force him and
Bob to collaborate. Then C collaborates with Eve, who tries to
eavesdrop on the line A-B in order to get as much as possible
information about Bob's qubit.

The security issue on these protocols is analogous to the two
partners case. We sketch the argument, see \cite{full} for all
details. The key of the demonstration is the fact that the time
ordering of the measurements is not important: if (say) the time
of Alice's measurement would change something in the local
statistics of her partners or in their correlations, the protocol
would allow signaling. Therefore, we can discuss security on
completely equivalent protocols in which some partners measure
their qubits first, this measurement acting as a preparation on
the state of the other qubits.

 Take Scenario 1 first: When Alice measures her
qubit, she prepares one of the four states
$\frac{1}{\sqrt{2}}(\ket{00}\pm\ket{11})$,
$\frac{1}{\sqrt{2}}(\ket{00}\pm i\ket{11})$. Therefore, one can
see this protocol as a two-partners communication, Alice sending
information to Bob-Charlie. Since we want to maximize $I(A:E)$ for
a fixed $I(A:BC)$, Eve's best individual attack can be copied
directly from (\ref{isometry}), replacing $\ket{0}_B$ and
$\ket{1}_B$ by $\ket{00}_{BC}$ and $\ket{11}_{BC}$ respectively:
\be
\begin{array}{lll}
U_{BCE}\ket{000}&=&\ket{000}\\
U_{BCE}\ket{110}&=&\cos\phi\ket{110}\,+\,\sin\phi\ket{001}
\end{array}\,.
\ee In particular, Eve can still perform the best individual
attack by using a single qubit. This is surprisingly simple,
because a priori Eve needs an increasing number of qubits
($2^{2n}$) to perform the most general attack on $n$ qubits.
However, even if Eve does not need a larger probe, she must be
able to implement a coherent operation on a bigger number of
qubits. Under this respect, eavesdropping on several channels is
more complicated than on a single channel.

Scenario 2 can be discussed in the same way: When A and C measure
their qubit, we have a single qubit flying to B, encoded as in the
BB84 protocol, and on which E eavesdrops. We just have to be
careful because the direct analogy with the two-partners case
gives us the optimum of $I(A:E)$ for a given value of $I(AC:B)$,
not of $I(A:BC)$. However, by the very definition of the protocol,
B and C are not correlated, whence $I(AC:B)=I(A:BC)$. Therefore,
in Scenario 2 Eve's best individual attack on Bob's qubit is
(\ref{isometry}).

\begin{figure}
\begin{center}
\epsfxsize=7cm \epsfbox{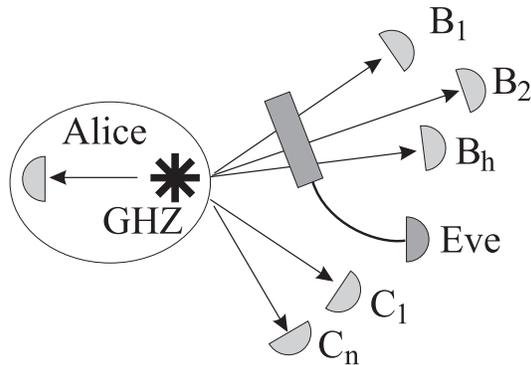} \caption{The general
eavesdropping scenario on N-QSS, with $h$ honest Bobs and
$n=N-h-1$ dishonest Charlies collaborating with Eve.}
\label{figqss}
\end{center}
\end{figure}

The same arguments can be worked out for the protocol N-QSS
involving $N$ partners. The general eavesdropping scenario is
shown in fig. \ref{figqss}: $n<N-1$ partners (Charlies,
$\underline{C}$) are dishonest, and want to retrieve the key
without the help of the $N-1-n\equiv h$ other partners (Bobs,
$\underline{B}$). Then again Eve can perform the best individual
attack using a single qubit, which must interact coherently with
all the $h$ qubits that are to be spied. The state of the $N+1$
qubits after eavesdropping is \ba
\ket{\Psi_{Nh}}&=&\frac{1}{\sqrt{2}}\big(
\ket{0^{N-h}}\ket{0^h}\ket{0}+\cos\phi
\ket{1^{N-h}}\ket{1^h}\ket{0}
\nonumber\\
&&+ \sin\phi \ket{1^{N-h}}\ket{0^h}\ket{1} \big)\label{psinh} \ea
where the first ket are A and the dishonest $\underline{C}$, the
second ket are the honest $\underline{B}$ that are spied, the
third ket is Eve. Let $I_a=I(A:\underline{B}\underline{C})$ the
information between the authorized partners,
$I_u=I(A:\underline{C}E)$ the information between the unauthorized
partners. In analogy with the case of two partners QKD, it can be
shown that $I_a>I_u$ if and only if $\phi<\frac{\pi}{4}$
\cite{full}. Now we can tackle the link with Bell's inequalities.

For our study, we consider the family of inequalities known as
Mermin-Klyshko (MK) inequalities \cite{belin,helle}. This choice
will be discussed below. The Bell operator for $M$ qubits is
defined recursively as \ba
B_M\,=\,\frac{\si_{a_M}+\si_{a_M'}}{2}\otimes B_{M-1}+
\frac{\si_{a_M}-\si_{a_M'}}{2}\otimes B_{M-1}'
\label{recurrence}\ea where $B_n'$ is obtained from $B_n$ by
exchanging all the $\vec{a}_k$ and $\vec{a}\,'_k$. The maximal
value allowed by QM is $S_M=2^{\frac{M+1}{2}}$, achieved for
$M$-qubit GHZ states. An important property of these inequalities
is the following: the bound $S_M\leq 2^{\frac{m+1}{2}}$, with
$m<M$, can be violated only by states in which more than $m$
qubits are entangled \cite{helle,refbell2}. We shall say that a
$M$-qubit state violates the inequality if for this state
$S_M>2^{\frac{M}{2}}$, that is, if the violation can be accounted
for only by having $M$-qubit entanglement.

Having settled these notions, we can prove\\
{\bf Theorem 2:} {\em The state $\ket{\Psi_{Nh}}$ given in
(\ref{psinh}) is such that the authorized partners violate the
$N$-qubit MK inequalitiy (in the sense just described) if and only
if $\phi<\frac{\pi}{4}$; and in this range, the unauthorized
partners do not violate the $(N-h+1)$-qubit MK inequality. At
$\phi=\frac{\pi}{4}$, both sets of partners are exactly at the
border of the violation; and for $\phi>\frac{\pi}{4}$ the roles of
the authorized and the unauthorized partners is reversed.}

The proof (see \cite{full} for all details) is a direct
optimization of expressions like
$\sandwich{\Psi_{Nh}}{B_N(\underline{a})\otimes\one}{\Psi_{Nh}}$
over all sets of $2N$ unit vectors $\underline{a}$. This
optimization is not easy. We could perform it analytically when
$N$ and $h$ have different parities (in particular, this is the
case if $h=N-1$, that is when all partners are honest and Eve is
external); and some cases where $N$ and $h$ have the same parity
were checked on the computer. Therefore, to within the limitations
of this proof, we can safely say that: for the N-QSS protocols,
and whatever the eavesdropping scenario in which Eve uses the best
individual attack, the security condition $I_a>I_u$ is satisfied
if and only if the authorized partners violate the MK inequality,
and in this case the unauthorized partners do not violate the MK
inequality. We recall that "violation" here does not merely mean
$S_M>2$, the limit imposed by lhv, but $S_M>2^{\frac{M}{2}}$, i.e.
that all the qubits are really strongly entangled.

One might ask if a purely algebraic result like Theorem 1 holds
for the violation of any $M$-qubit MK inequality. The answer is
negative. As a counterexample, the four-qubit state $\cos\alpha
(\ket{0011}+\ket{1100}+ i\ket{0101}+i\ket{1010})/2+\sin\alpha
(i\ket{1001}+ \ket{1111})/\sqrt{2}$ gives $S_{ABC}= S_{BCD}=
3\,>\,2\sqrt{2}$ for $\alpha\approx 0.955$. However, we have
numerical evidence that no such states can be produced by Eve. Our
current knowledge on this question can be found in \cite{full}. In
any case, the fact that a general algebraic theorem does not hold
in all cases {\em strengthens} the link between security and
violation of a MK inequality: even though in the Hilbert space we
can find states that violate two inequalities for some shared
qubits, these states do not appear in the individual eavesdropping
on a N-QSS protocol.

This leads us naturally to the question of the choice of the
optimal Bell's inequalities. Our choice of the MK inequalities is
natural in the following sense: we are considering QKD protocols
in which each partner measures two conjugated observables;
therefore, we choose also inequalities with two measurements per
qubit. Werner and Wolf have recently classified all the
inequalities of this class, and have demonstrated that the MK
inequalities are those that give the highest violation, for GHZ
states \cite{ww}. It is not impossible that other inequalities may
be better suited for the study of security in other protocols with
more than two settings per qubit. For instance, in the six-state
QKD between two partners it is known that $S_{AB}>2>S_{AE}$ is a
sufficient but not a necessary condition for security
\cite{revue,sixstate}.

Of course, we share the open questions of the whole field of
quantum cryptography: which is Eve's best attack in all
generality? Or, does something change if the partners share higher
dimensional systems instead of qubits? Note also that no
satisfactory Bell's inequality has been found yet for higher
dimensional systems. Under these respects, the study of the link
between Bell's inequalities and security seems to be a promising
field of research, at the border between quantum information and
foundations of quantum mechanics.

We conclude by stressing that Bell's inequalities appear here in a
context that is disconnected (at least at first sight) from the
studies on lhv: we have only discussed entanglement --- to be
precise, an entanglement that is "useful" for some quantum
communication protocols. In other words, Bell's inequalities seem
to have a role to play in "present-day" quantum information
processing, and not only in the "old" debate on lhv.

We acknowledge partial financial support from the Swiss FNRS and
the Swiss OFES within the European project EQUIP (IST-1999-11053).


\begin{thebibliography}{10}
\bibitem{notef} Writing $\vec{a}_{+}=\vec{a}$, $\vec{a}_{-}=\vec{a}\,'$,
$\vec{A}_{+}=\vec{A}$ and $\vec{A}_{-}=\vec{A}\,'$, the full
expression for $f$ is \ban
f&=&\big|\vec{a}\wedge\vec{a}\,'\big|^2\,
\big|\vec{b}\wedge\vec{b}\,'\big|^2\,+\,\big|\vec{A}\wedge\vec{A}\,'\big|^2
\big|\vec{c}\wedge\vec{c}\,'\big|^2\,+\\
&&+\,\sum_{\si,\omega=\pm}
(\vec{a}_{\si}\cdot\vec{A}_{\omega})^2(1+\si\vec{b}\cdot\vec{b}\,')
(1+\omega\vec{c}\cdot\vec{c}\,')\,.\ean
\bibitem{note1} The same behavior is shown by the states
introduced in \cite{bruno}, where it was erroneously claimed that
both pairs Alice-Bob1 and Alice-Bob2 violate the CHSH inequality.
\end{thebibliography}
\end{document}